\documentstyle[12pt,epsf]{article}
\newcommand{\bea}{\begin{eqnarray}}
\newcommand{\eea}{\end{eqnarray}}
\newcommand{\be}{\begin{equation}}
\newcommand{\ee}{\end{equation}}

\begin{document}
\hfill
$\vcenter{
\hbox{\bf LC-TH-2000-006}
}$
\medskip
\begin{Large}
\begin{center}
{\bf $Z^\prime$ indication from new APV data in Cesium
and searches at linear colliders}
\end{center}
\end{Large}
\vskip 1.0cm
\centerline {R. Casalbuoni$\;^{a,b)}$,
S. De Curtis$\;^{b)}$}
\centerline {D. Dominici$\;^{a,b)}$, R. Gatto$\;^{c)}$
and S. Riemann$\;^{d)}$}
\vskip 0.5cm

\noindent
$a)$ {\it Dipartimento di Fisica, Univ. di Firenze, I-50125 Firenze, Italia.}
\hfill\break\noindent
$b)$ {\it I.N.F.N., Sezione di Firenze, I-50125 Firenze, Italia.}
\hfill\break\noindent
$c)$ {\it D\'ept. de Phys. Th\'eorique, Univ. de Gen\`eve, CH-1211 Gen\`eve
4, Suisse.}
$d)$ {\it DESY Zeuthen, Platanenallee 6, D-15738 Zeuthen, Germany.}
\vskip 1cm
\section{Introduction}
In a recent paper \cite{bennett} a new determination of the weak
charge of atomic cesium has been reported.
The most precise atomic parity violating (APV) experiment compares the
mixing among $S$ and $P$ states due to neutral weak interactions
to an induced Stark mixing \cite{wood}. The 1.2\% uncertainty
on the previous measurement of the
weak charge $Q_W$ was dominated by the
theoretical
calculations on the amount of Stark mixing  and on the
electronic parity violating
 matrix elements. In  \cite{bennett}
the Stark mixing was measured and, incorporating new experimental
data, the uncertainty in the electronic  parity violating matrix elements
was reduced.
The new result
\be
Q_W(^{133}_{55}Cs)=-72.06\pm (0.28)_{\rm expt}\pm (0.34)_{\rm
theor}
\label{newexp}
\ee
represents a considerable improvement with respect to the previous determination
\cite{wood,noecker,blundell}
\be
Q_W(^{133}_{55}Cs)=-71.04\pm (1.58)_{\rm expt}\pm (0.88)_{\rm
theor}
\ee
On the theoretical side, $Q_W$ can be expressed in terms of the
$S$ parameter  \cite{marciano} or the $\epsilon_3$
\cite{altarelliqw}
\be
Q_W=-72.72\pm 0.13-102\epsilon_3^{\rm rad}+\delta_NQ_W
\ee
including hadronic-loop uncertainty. We use here the variables
$\epsilon_i$ (i=1,2,3) of ref. \cite{altarelli}, which include the
radiative corrections, in place of the set of
variables $S$, $T$ and $U$ originally introduced in ref. \cite{peskin}.
In the above
definition of $Q_W$
we have  explicitly included only the Standard Model (SM) contribution
to the radiative corrections. New physics (that is physics beyond the
SM) is represented by the term $\delta_NQ_W$
including also contributions
to $\epsilon_3$. Also, we
have neglected a correction proportional to $\epsilon_1^{\rm rad}$.
In fact, as well known \cite{marciano}, due to the
particular values of the number of neutrons ($N=78$) and of
protons ($Z=55$) in cesium, the dependence on
$\epsilon_1$ almost cancels out.

From the theoretical expression
we see that $Q_W$ is particularly sensitive to new physics
contributing to the $\epsilon_3$ parameter. This kind of new
physics is severely constrained by the high energy experiments.
From a recent fit  \cite{altarelli2},
the  value of $\epsilon_3$ from the  high energy data is
$\epsilon_3^{\rm expt}=(4.19\pm 1.0)\times 10^{-3}$.

To estimate new physics contributions to this parameter one
has to subtract the SM radiative corrections, which, for
$m_{top}=175~GeV$ and for $m_H~(GeV)=100,~300$, are given
respectively by
\be
\matrix{m_H=100~GeV && \epsilon_3^{\rm rad}=5.110\times 10^{-3}\cr
        m_H=300~GeV && \epsilon_3^{\rm rad}=6.115\times
        10^{-3}}
\ee
Therefore new physics contributing to $\epsilon_3$ cannot be larger than a few
per mill. Since $\epsilon_3$ appears in $Q_W$ multiplied by a factor 102, the
kind of new physics which contributes through $\epsilon_3$ cannot contribute to
$Q_W$ for more than a few tenth. On the other side the discrepancy between the
SM and the experimental data is given by (for a light Higgs)
\be
Q_W^{\rm expt}-Q_W^{SM}=1.18\pm 0.46
\ee
where we have added in quadrature the uncertainties.
This  corresponds to 2.6(2.8)-$\sigma$ deviation with respect to the SM for
$m_H=100(300)~GeV$.
The 95\%CL limits
on $\delta_NQ_W$ are
\be
0.28\leq \delta_NQ_W\leq 2.08~~{\rm for}~~m_H=100~GeV
\label{bounds}
\ee
\be
0.38\leq \delta_NQ_W\leq 2.18~~{\rm for}~~m_H=300~GeV\\
\label{bounds2}
\ee
For increasing $m_H$ both bounds increase.
One possible contribution to $Q_W$ which was neglected is the difference
between neutron and proton spatial distributions in the nucleus. With the
increasing  APV measurement precision, this effect has been reconsidered
\cite{pollock} adding an additional
$\Delta Q^{n-p}_W =\pm 0.3$ to the theoretical
error of eq. (\ref{newexp}). Including this additional uncertainty, the
theoretical error in eq. (\ref{newexp}) becomes $\pm (0.45)$ and the
deviation with respect to SM for $m_H=100~GeV$ decreases from 2.6-$\sigma$
to 2.2-$\sigma$. Our analysis will be based on the result given
in eq. (\ref{newexp}).
The lower positive bounds from eqs. (\ref{bounds}) and  (\ref{bounds2}) exclude the
SM at 99\%CL  and, a fortiori, all the models leading to negative extra
contribution to $Q_W$, as for example models with a sequential $Z^\prime$
\cite{apv}. This 2.6(2.8)-$\sigma$ deviation with respect to the SM
for $ m_H=100(300)~GeV$ could be explained by assuming the existence of an extra
$Z^\prime$ from $E_6$ or $O(10)$ or from $Z^\prime_{LR}$ of left-right (LR) models
\cite{bennett,apv,rosner,erlanga}.

\section{Bounds on extra $Z^\prime$ from $Q_W$}

Let us now look at models which, at least in principle, could
give rise to a sizeable modification of $Q_W$. In ref.
\cite{altarelli3} it was pointed out that models involving extra
neutral vector bosons coupled to ordinary fermions can do the
job.
The high energy data at the $Z$ resonance strongly bound the $Z-Z'$ mixing
\cite{gross}.
 For this reason we will assume zero mixing  in the following calculations. In
this case $\delta_NQ_W$ is due to the direct exchange of the $Z^\prime$ and is
completely fixed by the $Z^\prime$ parameters:
\be
\delta_NQ_W=
16 a_e^\prime [(2Z+N)
v_u^\prime+
(Z+2N)  v_d^\prime]
\frac{M_Z^2}{M_{Z^\prime}^2}
\label{deltaQ}
\ee
$a_f^\prime,
v_f^\prime$ are the couplings  $Z^\prime$ to fermions.

We will discuss the following classes of models involving an extra $Z^\prime$:
 the LR  models and the extra-U(1) models. The relevant couplings of
 the $Z^\prime$ to the electron and to the up and down quarks are given in the
Table 1 of \cite{apv}.
 The different extra-U(1) models are parameterized by the angle $\theta_6$. In
\cite{apv} we used a different definition of the angle. The relation between the
angle $\theta_2$ of ref. \cite{apv} and $\theta_6$ is given by
$\theta_6=\theta_2- \arctan \sqrt{5/3}$.
\begin{figure}[t]
\epsfysize=8truecm
\centerline{\epsffile{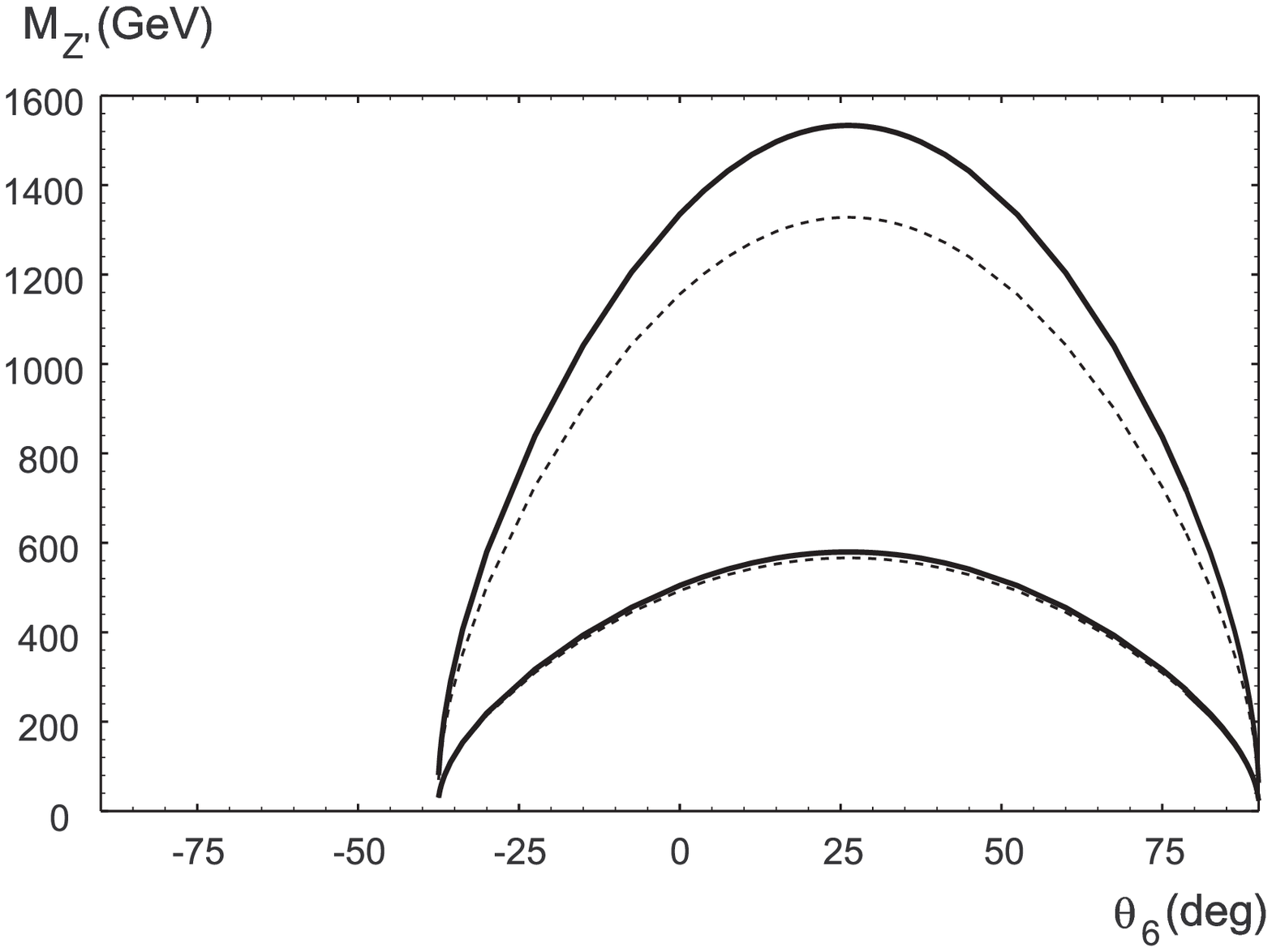}}
{\bf Figure 1} - {\it The 95\%CL lower and upper bounds for $M_{Z^\prime}$ for the
extra-U(1) models versus $\theta_6$. The solid and the dash lines
correspond to $m_H=100~GeV$ and $m_H=300~GeV$ respectively. The lower bounds
from direct search at Tevatron  is about $600 ~GeV$.}
\label{apvz}
\end{figure}

In the case of the  LR model  considered in \cite{apv}
the extra contribution to the weak charge is
\be
\delta_NQ_W=-\frac{M_Z^2}{M_{Z^\prime}^2}Q_W^{SM}~~~~
\ee
For this model one has a 95\%CL lower bound on $M_{Z^\prime_{LR}}$ from Tevatron
\cite{tevatron} given by $M_{Z^\prime_{LR}}\ge 630~GeV$. A LR model could then
explain the APV data allowing for a mass of the $Z^\prime_{LR}$ varying between
the intersection from the 95\%CL bounds $540\le M_{Z^\prime_{LR}}(GeV)\le 1470$
deriving from eq. (\ref{bounds}) and the lower bound of $630~GeV$ .

In the case
of the extra-U(1) models  the CDF experimental lower bounds for the masses vary
according to the values of the parameter $\theta_6$ which parameterizes
different extra-U(1) models, but in general they are about $600~ GeV$ at 95\%CL
\cite{tevatron}.
  In particular for the model known in the literature as  $\eta$ (or $A$), which
 corresponds to $\theta_6=\arctan{-\sqrt{5/3}}$, the 95\%CL lower bound is
 $M_{Z^\prime_\eta}\simeq 620~GeV$, for the model
  $\psi$ (or  $C$), which corresponds to $\theta_6=\pi/2$, the lower bound is
$M_{Z^\prime_\psi}\simeq 590~GeV$ and for the model $\chi$, which corresponds to
$\theta_6=0$, the lower bound is $M_{Z^\prime_\chi}\simeq 595~GeV$.

By comparing eqs. (\ref{bounds}), (\ref{bounds2}) with eq.   (\ref{deltaQ}) we
see that the models $\eta$ and $\psi$ are excluded. The bounds on $\delta_N Q_W$
at 95\%CL  can be translated into  lower and upper bounds on $M_{Z^\prime}$. The
result is given in Fig. 1,
where the bounds are plotted versus $\theta_6$. In
looking at this figure one should also remember that the direct lower bound from
Tevatron is about $600 ~GeV$ at 95\%CL. We see that the presence of an extra
$Z^\prime$ can explain the discrepancy with the SM prediction for the $Q_W$ for
a wide range of $\theta_6$ angle.  In particular  the $\chi$ (or $B$) model,
corresponding to  $\theta_6=0$, is  allowed for $M_{Z^\prime_\chi}$ less than
about 1.2 $TeV$.
\begin{figure}[t]
\begin{center}
\epsfysize=9truecm
\centerline{\epsffile{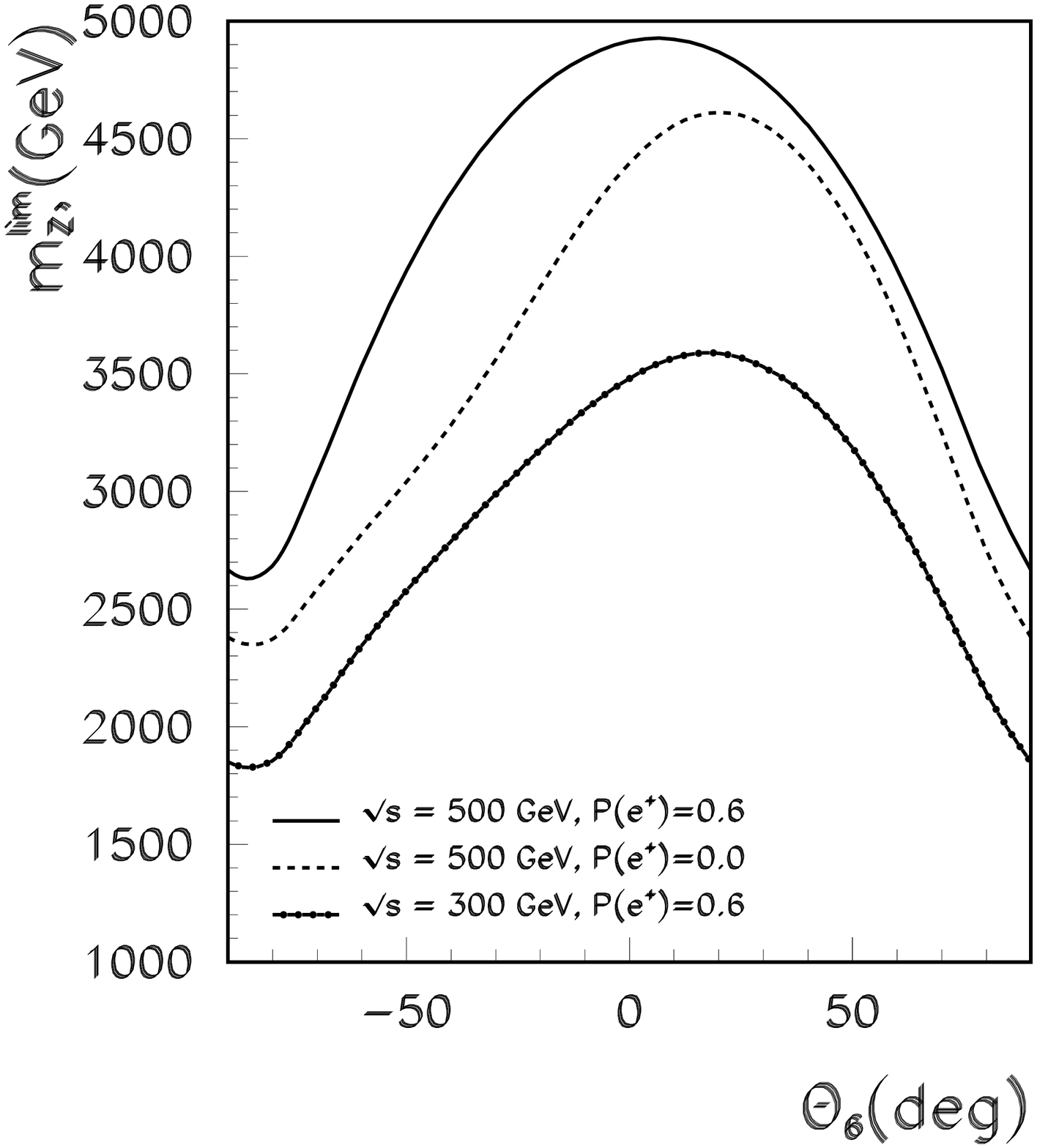}}
\end{center}
{\bf Figure 2} - {\it 95\%CL lower bounds for $M_{Z^\prime}$ for the extra-U(1)
models versus $\theta_6$ from a LC with $\sqrt{s}=500~GeV$,
$L=500~fb^{-1}$, $P_{e^-}=0.9$, $P_{e^+}=0.6$  (solid line),
 $P_{e^-}=0$ (dash line), $\sqrt{s}=300~GeV$,
$L=300~fb^{-1}$, $P_{e^-}=0.9$, $P_{e^+}=0.6$  (solid-dot line).
}
\end{figure}
\section{$Z^\prime$ at future colliders}
 The search for a $Z^\prime$ is one of the tasks of future colliders. The
 existing bounds for $E_6$ models, $M_{Z^\prime}\sim 600~GeV$ from direct search
at Tevatron will be upgraded by the future run with $\sqrt{s}=2~TeV$,
$L=1~fb^{-1}$ to $M_{Z^\prime}\sim 800-900~GeV$ and pushed to $\sim 1~TeV$ for
$L=10~fb^{-1}$. The bounds are based on 10 events in the
$e^+e^-+\mu^+
\mu^-$ channels and decays to SM final states only is assumed \cite{rizzo}.
At the LHC with an integrated luminosity of $100~fb^{-1}$ one can explore a mass
range up to $4-4.5~TeV$ depending on the $\theta_6$ value.
 Concerning LR models, the 95\%CL lower limits from Tevatron run with
$\sqrt{s}=2~TeV$, $L=1(10)~fb^{-1}$ are $\sim 900(1000)~GeV$ and extend to $\sim
4.5 ~TeV$ at LHC \cite{rizzo}.
\begin{figure}[t]
\begin{center}
\epsfysize=9truecm
\centerline{\epsffile{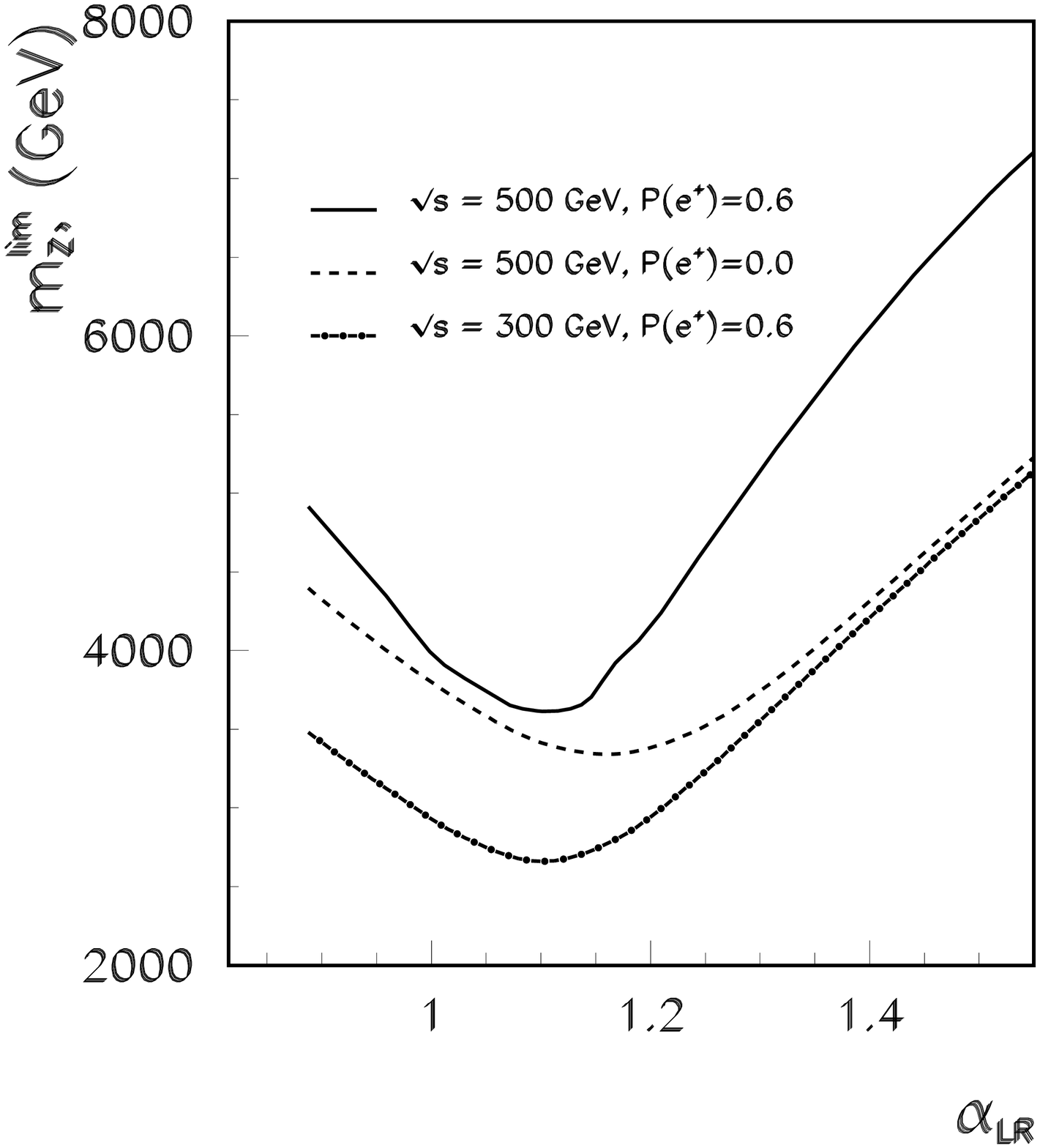}}
\end{center}
{\bf Figure 3} - {\it Same of Fig. 2
 for $M_{Z'_{LR}}$ for the LR
models versus  $\alpha_{LR}$.
}
\end{figure}
Therefore if the deviation for $Q_W$ with respect to the SM prediction is not
due to a statistical fluctuation but to the presence of new physics like new
extra gauge bosons from $E_6$ or LR models, LHC can verify or disprove
this possible evidence. However little can be learned on the $Z^\prime$
properties.
 With $e^+e^-$ colliders the properties of a $Z^\prime$ can be easily
 investigated if the center-of-mass energy is large enough to produce it.
 Anyway, from the measurements of $e^+e^-\to\gamma,Z,Z^\prime\to\bar f f $ below
 threshold one can get information about the nature of the $Z^\prime$.
\begin{figure}[t]
\begin{center}
\epsfysize=9truecm
\centerline{\epsffile{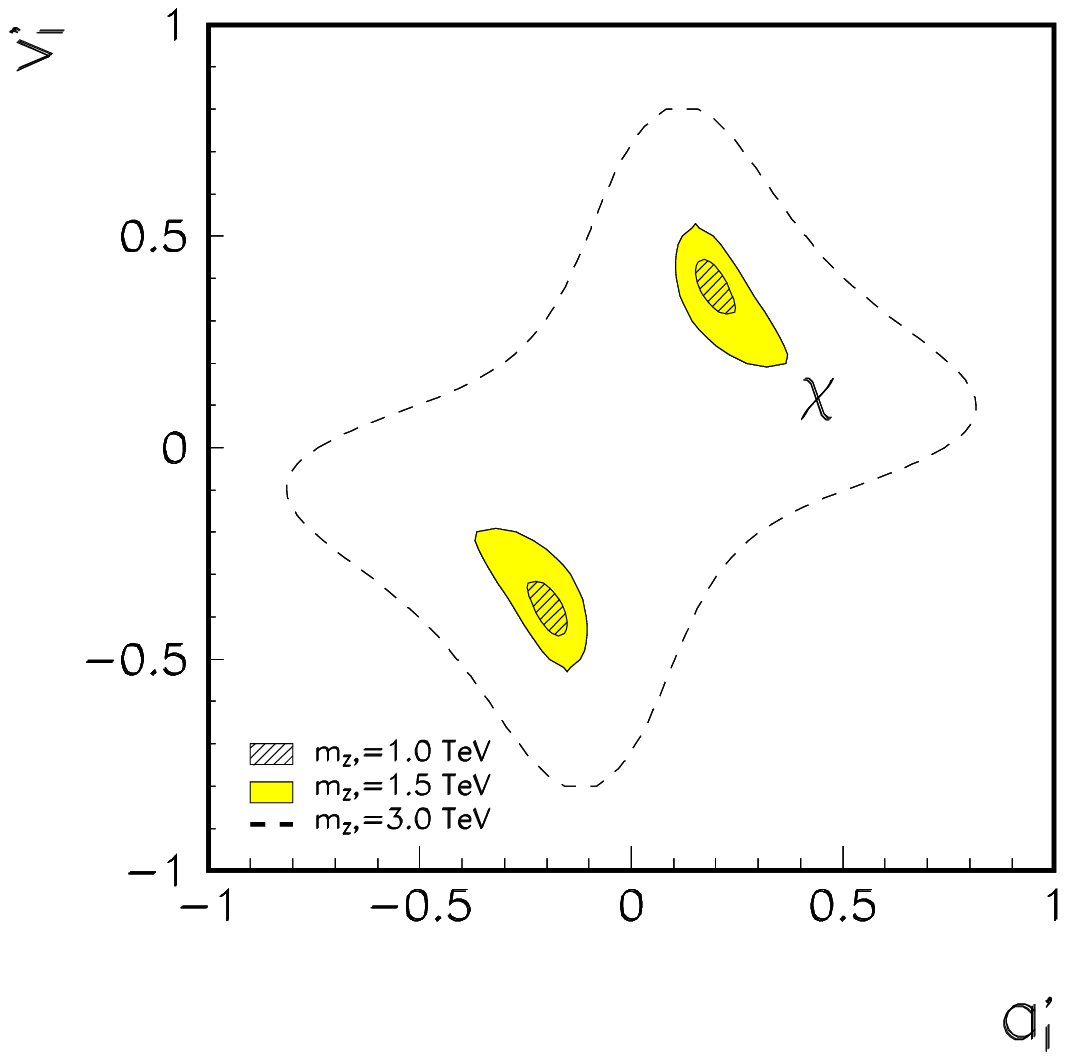}}
\end{center}
{\bf Figure 4} - {\it 95\%CL contours for
$(a_l^\prime$, $v_l^\prime)$
for  $M_{Z^\prime}=1,1.5,3~TeV$, and
$\sqrt{s}=300~GeV$ and $L=300~fb^{-1}$.
}
\end{figure}
\begin{figure}[t]
\begin{center}
\epsfysize=9truecm
\centerline{\epsffile{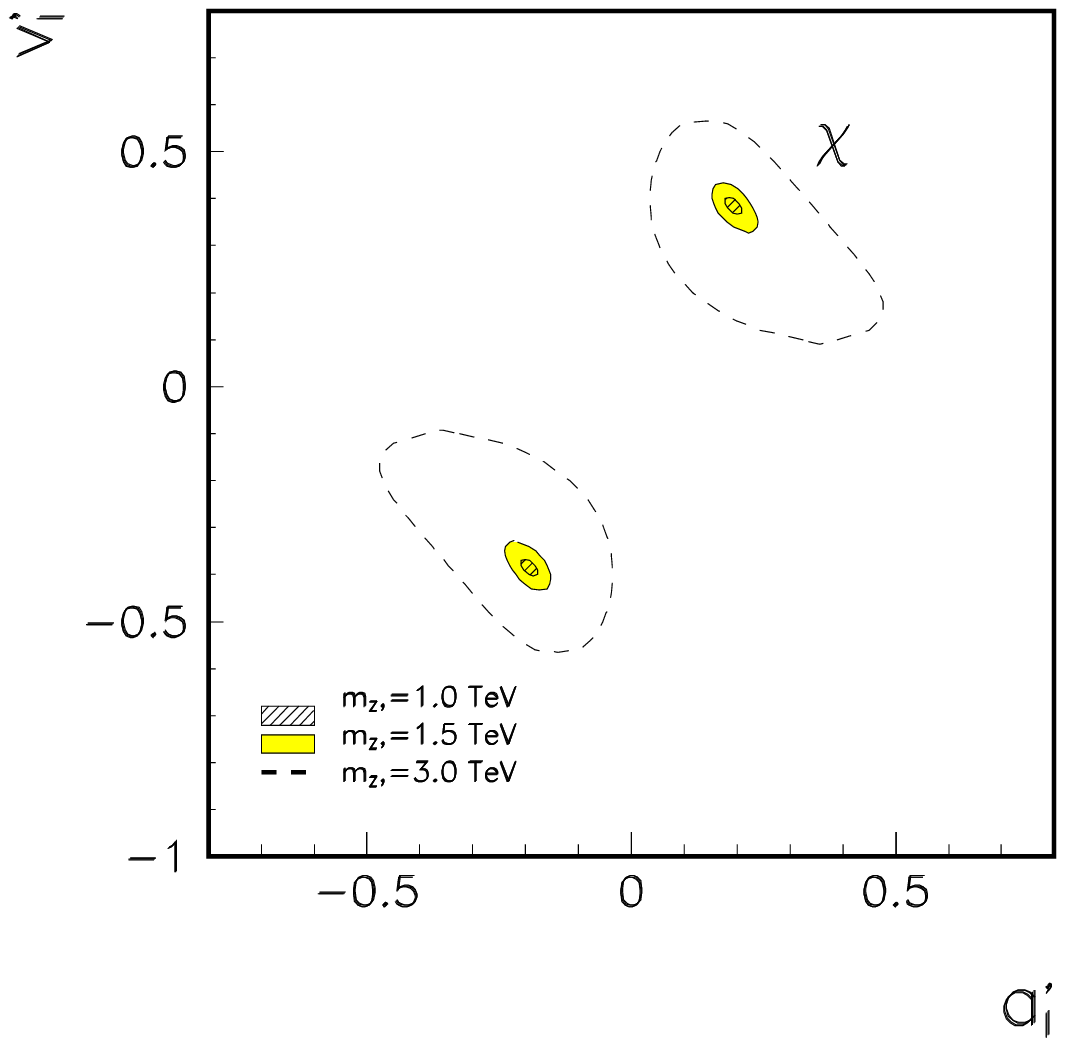}}
\end{center}
{\bf Figure 5} - {\it Same of Fig. 4 for
$\sqrt{s}=500~GeV$ and $L=500~fb^{-1}$.
}
\end{figure}
In ref.
 \cite{sabine}  exclusion limits for $M_Z^\prime$ have been determined
 from the analysis of all leptonic and hadronic observables. Here we
 show the upgrading of this analysis for a collider scenario
$\sqrt{s}=500~GeV$, $L=500~fb^{-1}$ and polarized beams: $P_{e^-}=0.9$, $P_{e^+}=0.6$
(solid line).
The  95\%CL lower bounds on $M_{Z^\prime}$ are given  in
 Fig. 2
 for different values of $\theta_6$ for $E_6$ models. Also shown
is the case of unpolarized positron beam (dash line) and the case of
$\sqrt{s}=300~GeV$, $L=300~fb^{-1}$ (solid-dot line).
 The observables which are considered are the total lepton and hadron
cross sections,
 the forward backward asymmetry, the left-right asymmetry
(for leptons, hadrons, $c\bar c$  and
$b\bar b$ final states), $ R_b$ and $R_c$.
 The assumed identification efficiencies are: for leptons $\epsilon_l=95\%$,
for
 $c\bar c$  $\epsilon_c=40 \%$, for $b\bar b$
 $\epsilon_b=60\%$
and the
corresponding systematic errors $\Delta \epsilon_l/\epsilon_l=0.5\%$
$ \Delta \epsilon_c/\epsilon_c= 1.5\%$, and
 $\Delta\epsilon_b/\epsilon_b=1\%$.
An uncertainty of 0.5\% on the luminosity
and $\Delta P/P=1\%$ are considered.

 The same analysis
 for the LR models  leads to a 95\%CL lower
 bound for $M_{Z^\prime_{LR}}$ shown in  Fig. 3.
 For example the particular LR
model considered in  \cite{apv} corresponds to
$\alpha_{LR}=\sqrt{\cot^2\theta_W-1}$ and for this
model the bound extracted from Fig. 3 is around $7~ TeV$.

Assuming that the $Z^\prime$ mass is known, one can study the
$Z^\prime$ couplings to fermions.
 The 95\%CL contours on $Z^\prime \bar ll$
couplings  for the $\chi$ model
for $M_{Z^\prime}=1,1.5,3~TeV$ are presented
in Fig. 4 for $\sqrt{s}=300~GeV$ and $L=300~fb^{-1}$
and in Fig. 5 for $\sqrt{s}=500~GeV$ and $L=500~fb^{-1}$.
\begin{figure}[t]
\begin{center}
\epsfysize=9truecm
\centerline{\epsffile{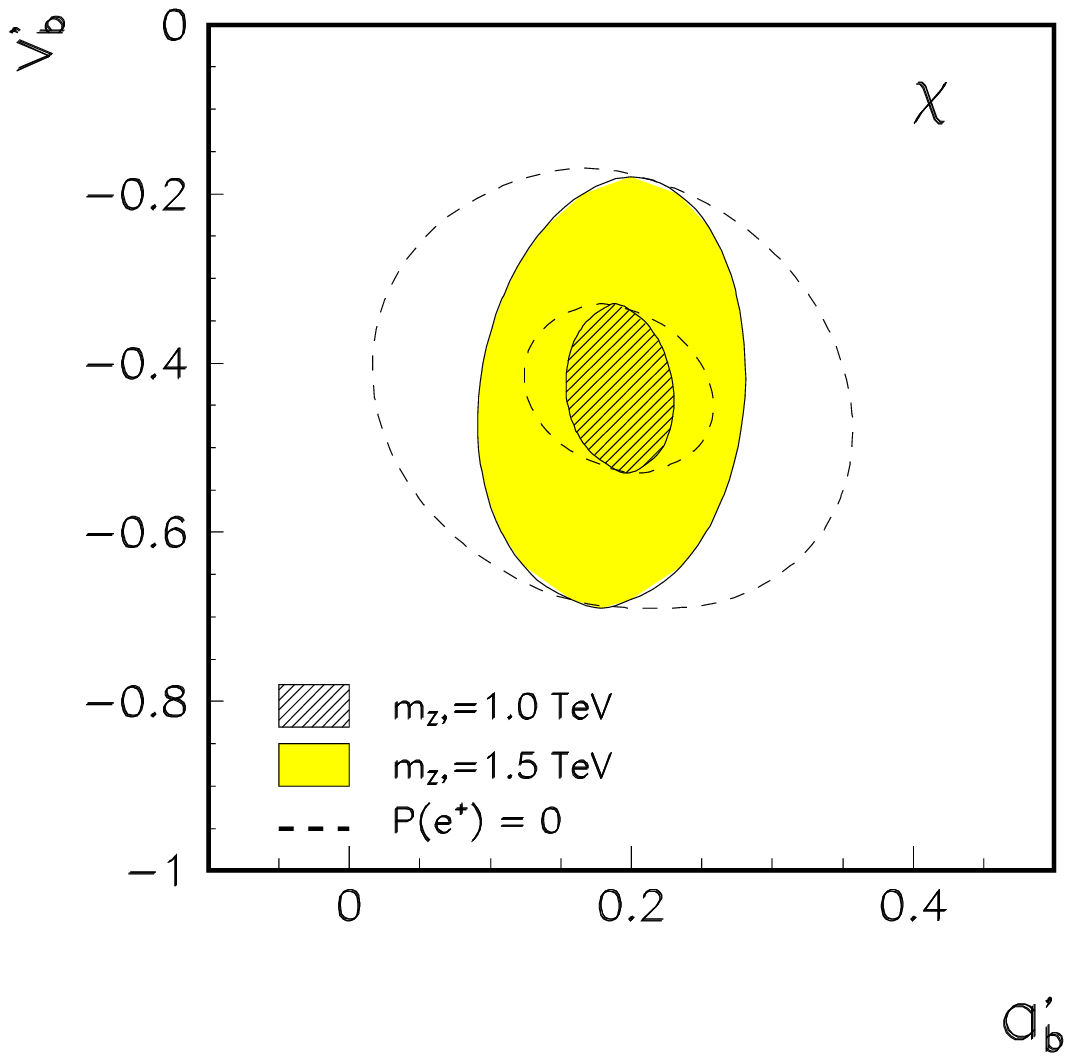}}
\end{center}
{\bf Figure 6} - {\it 95\%CL contours for
$(a_b^\prime$, $v_b^\prime)$
for  $M_{Z^\prime}=1,1.5~TeV$, and
$\sqrt{s}=500~GeV$ and $L=500~fb^{-1}$.
The dash lines correspond to
$P_{e^+}=0$.
}
\end{figure}
In Fig. 6
the 95\%CL contours for $a_b^\prime$ $v_b^\prime$
for  $M_{Z^\prime}=1,1.5~TeV$, and
$\sqrt{s}=500~GeV$ and $L=500~fb^{-1}$
are shown.
The dash lines correspond to
$P_{e^+}=0$.

In conclusion, once a $Z^\prime_\chi$ is seen at LHC, already
a LC with $\sqrt{s}=300~GeV$, $L=300~fb^{-1}$ and polarized beams
is able to measure the $Z^\prime_\chi$ couplings to fermions with
good precision. For example for a $Z^\prime_\chi$ with mass of $1~ TeV$
(which can explain the new APV result) the couplings to leptons can be determined
within 10\%, unless a sign ambiguity. By increasing the center of mass energy, the
precision improves.


\begin{thebibliography}{9}


\bibitem{bennett}
S.C. Bennett and C.E. Wieman, {\it
Phys. Rev. Lett.} {\bf 82} (1999) 2484

\bibitem{wood}
C.S. Wood  et al., {\it  Science} {\bf 275} (1997) 1759

\bibitem{noecker}
M.C. Noecker, B.P. Masterson and C.E. Wieman, {\it
 Phys. Rev. Lett.} {\bf 61}  (1988) 310

\bibitem{blundell}
S.A. Blundell, W.R. Johnson and J. Sapirstein, {\it Phys. Rev. Lett.}
{\bf 65} (1990) 1411; V. Dzuba, V. Flambaum,  P. Silvestrov and
O. Sushkov, {\it Phys. Lett.} {\bf A141} (1989)  147
\bibitem{marciano}
W.J. Marciano  and  J.L. Rosner, {\it Phys. Rev. Lett.} {\bf 65} (1990)
2963

\bibitem{altarelliqw}
See the second reference in \cite{altarelli}

\bibitem{altarelli}
G. Altarelli, R. Barbieri and  S. Jadach, {\it Nucl. Phys.} {\bf B369} (1992)
 3; G. Altarelli, R. Barbieri and F. Caravaglios, {\it Nucl.
Phys.} {\bf B405} (1993) 3; {\it ibidem},  {\it Phys. Lett.} {\bf B349} (1995)
145
\bibitem{peskin}
M.E. Peskin  and T. Takeuchi, {\it Phys. Rev. Lett.} {\bf 65} (1990) 964
and  {\it Phys. Rev.} {\bf D46} (1991) 381
\bibitem{altarelli2}
G. Altarelli, private communication

\bibitem{pollock}
S.J. Pollock and  M.C. Welliver, {\it Phys. Lett.} {\bf B464} (1999) 177

\bibitem{apv}
R. Casalbuoni, S. De Curtis,
 D. Dominici  and R. Gatto, {\it Phys. Lett.} {\bf B460} (1999) 135

\bibitem{rosner} J.L. Rosner, hep-ph/9907524

\bibitem{erlanga}
J. Erler and  P. Langacker, hep-ph/9910315

\bibitem{altarelli3}
G. Altarelli, R. Casalbuoni, S. De Curtis, N. Di Bartolomeo,
F. Feruglio and  R. Gatto, {\it Phys. Lett.} {\bf B261} (1991) 146

\bibitem{gross}
E. Gross, Contribution to the International Europhysics
Conf., Tampere, Finland, 15-21 July


\bibitem{tevatron}
F. Abe et al., {\it Phys. Rev. Lett.}  {\bf 79} (1997) 2191

\bibitem{rizzo}
T.G. Rizzo, in "New Directions For High-Energy Physics", Eds. D.G. Cassel,
L. Trindle Gennari, R.H. Siemann, Stanford, CA,
 (1997) and references therein

\bibitem{sabine} S. Riemann,
 in "$e^+e^-$ Linear Colliders: Physics and Detector Studies",
 Ed. R. Settles, DESY 97-123E, p.227 (1997).


\end{thebibliography}
\end{document}